# Event Detection and Localization in Distribution Grids with Phasor Measurement Units

Omid Ardakanian*, Ye Yuan†, Roel Dobbe‡, Alexandra von Meier‡, Steven Low§ and Claire Tomlin‡
*Electrical and Computer Engineering Department, University of British Columbia
†School of Automation, Huazhong University of Science and Technology, for correspondence: yye@hust.edu.cn
‡Electrical Engineering and Computer Sciences Department, University of California at Berkeley
§Computing and Mathematical Science Department, California Institute of Technology

*Abstract*—The recent introduction of synchrophasor technology into power distribution systems has given impetus to various monitoring, diagnostic, and control applications, such as system identification and event detection, which are crucial for restoring service, preventing outages, and managing equipment health. Drawing on the existing framework for inferring topology and admittances of a power network from voltage and current phasor measurements, this paper proposes an online algorithm for event detection and localization in unbalanced three-phase distribution systems. Using a convex relaxation and a matrix partitioning technique, the proposed algorithm is capable of identifying topology changes and attributing them to specific categories of events. The performance of this algorithm is evaluated on a standard test distribution feeder with synthesized loads, and it is shown that a tripped line can be detected and localized in an accurate and timely fashion, highlighting its potential for real-world applications.

## I. INTRODUCTION

Increased adoption of distributed energy resources (DER) in recent years has led to an unprecedented level of variability and uncertainty in distribution systems, creating new challenges in maintaining safe and reliable operation, and increasing the resilience of the grid. Feeder behavior is increasingly harder to predict and new protection issues arise, such as desensitization and unintended islanding or tripping [1]. This can lead to accelerated structural damage and potentially cascading failures, and yield economic burden due to accelerated wear [2]. The rapid adoption of electric vehicles (EVs) will further aggravate this situation [3], especially if charging is optimized for electricity prices [4]. In addition, the inability to assess the impact of DER and EVs on the network leads utilities to impose conservative caps on the allowable DER capacity and number of EVs, hindering the transition to a renewable energy infrastructure. These concerns have mobilized many distribution systems operators (DSOs) to build a stronger information layer on top of the physical infrastructure exploiting recent advances in sensing and communication.

Traditionally, DSOs had little need for monitoring and diagnostic capabilities and relied on field personnel to report the status of network equipment in order to determine the topology at a specific time. Moreover, outages and other critical events would remain undetected unless they are reported by the customers or proliferated to the level of a manned substation. Recently, many US DSOs have begun deploying high-precision distribution phasor measurement units (PMUs) for monitoring, diagnostic, and control purposes [5]. High resolution voltage and current phasor measurements can be used in a plethora of applications concerning real-time system operation and long-term planning [6], such as state estimation [7], model validation, load characterization, and *event detection and localization* which is the focus of this work. Event detection is the problem of detecting the occurrence of safety-critical events in a power system, such as outages, switching operations, or cyber attacks, while event localization deals with attributing such events to specific network components and a small geographic area of the network. Accurate and timely detection and localization is crucial for determining remedial control actions in order to prevent cascading outages, restore service, and manage the health of critical equipment. However, developing event detection and identification techniques can be quite challenging since PMU coverage is often limited at the distribution level, i.e., many buses are not monitored [5]. Moreover, voltage measurements in three-phase distribution systems tend to be spatially coupled leading to low dimensionality in the setting of network inference [8]. Lastly, PMU readings are subject to noise [9], which can corrupt an inference algorithm.

This paper builds on our prior work on the inverse power flow (IPF) problem [10] which concerns inferring the admittance matrix of a radial or a mesh network from measured voltage and current phasors of all buses or just a subset of them. We extend the IPF framework to a *three-phase* distribution system and develop an online algorithm for event detection and localization, which tackles the low rank structure of the PMU data using convex relaxation and matrix partitioning techniques. The proposed algorithm does not require a priori knowledge of the underlying network topology and relies on PMU data only. We show that any event that induces a change in the admittance matrix can be detected immediately after it occurs and its type and approximate location can also be determined in a sub-second time frame[1]. Simulations performed on the IEEE 13 bus test feeder confirm that the proposed algorithm is effective in identifying a tripping event using a small number of PMU samples and can quickly recover an important part of the admittance matrix from high-precision voltage and current phasor measurements.

---

[1]An existing synchrophasor technology for distribution grids, termed $\mu$PMU, samples AC voltage and current waveforms at 2 samples per cycle [11]. Hence, an event can be detected within a few hundred milliseconds of its occurrence using an algorithm that requires a small number of samples.

## II. RELATED WORK

With the recent availability of massive PMU data from transmission and distribution systems, a growing body of research has been built on developing algorithms for network topology identification [12]–[14], and event detection [8], [15]–[22]. In particular, several algorithms are proposed for event detection in distribution systems. For example, Cavraro et al. [22] propose a data-driven online algorithm for detecting a switching event that changes the topology of a distribution network by comparing a trend vector built from PMU data with a given library of signatures derived from possible topology changes. This algorithm cannot be applied to detect other events since obtaining the signature of all possible events is impractical. Sharon et al. [21] investigate the optimal placement of sensors in a distribution network in order to infer the status of switches from their measurements with high confidence using a maximum likelihood method. This approach requires knowledge of the number of switches installed in the network and their location and cannot be extended to other types of events. The closest line of work to ours is by Xie et al. [8] which uses principal component analysis to obtain a lower dimensional subspace of the available PMU data, and projects the original data onto this subspace by learning coefficients of the basis matrix using an adaptive training method. An online event detection algorithm is then proposed to approximate PMU measurements using these coefficients, issuing an alert whenever a significant approximation error is noticed. This work merely focuses on event detection and does not investigate the localization problem.

In short, to the authors' knowledge, the current literature focuses on detecting specific types of events without addressing event localization and classification problems. The approach in this paper fills that gap as it both detects and determines the approximate location of any event that induces a change in the admittance matrix. Building on our prior work [10], we extend the method from single- to three-phase AC power flow models, taking into account the coupling between phases.

## III. PROBLEM FORMULATION

In this section we formulate the three-phase version of the IPF problem presented in [10] and propose a novel algorithm for identifying the admittance matrix which deals with the low rank structure of the PMU data.

### A. Three-Phase Extension of the IPF Problem

Let $\mathbb{C}$ denote the set of complex numbers and $\mathbb{S}$ denote the set of symmetric complex matrices. For $A \in \mathbb{C}^{n \times n}$, let $\operatorname{Re} A$ and $\operatorname{Im} A$ denote matrices with the real and imaginary parts of $A$, respectively. The transpose of a matrix $A$ is denoted $A^\top$, its Hermitian (complex conjugate) transpose is denoted $A^H$, and its pseudo-inverse is denoted $A^\dagger$.

A three-phase power distribution system can be modeled by an undirected graph $\mathcal{G} = (\mathcal{N}, \mathcal{E})$ where $\mathcal{N} = \{1, 2, \ldots, N\}$ represents the set of nodes, and $\mathcal{E} \subseteq \mathcal{N} \times \mathcal{N}$ represents the set of overhead or underground lines, each connecting two distinct nodes. We denote the phases of a node $n \in \mathcal{N}$ by $\mathcal{P}_n \subseteq \{a_n, b_n, c_n\}$ and the phases of a line $(m, n) \in \mathcal{E}$ by $\mathcal{P}_{mn} \subseteq \{a_{mn}, b_{mn}, c_{mn}\}$. Let $V_n^\phi \in \mathbb{C}$ be the line-to-ground voltage at node $n \in \mathcal{N}$ of phase $\phi \in \mathcal{P}_n$ and $I_n^\phi \in \mathbb{C}$ be the current injected at the same node and phase. We denote voltages and injected currents of phases at node $n \in \mathcal{N}$ by vectors $V_n = [V_n^\phi|_{\phi \in \mathcal{P}_n}]^\top$ and $I_n = [I_n^\phi|_{\phi \in \mathcal{P}_n}]^\top$, respectively. Let $D = \sum_{n \in \mathcal{N}} |\mathcal{P}_n|$ be the number of node/phase pairs in the network. Assuming that node 1 represents the distribution substation, we treat $V_1$ as reference for phasor representation.

We model lines as $\pi$-equivalent components and denote the phase impedance and shunt admittance matrices of line $(m, n)$ by $Z_{mn} \in \mathbb{C}^{|\mathcal{P}_{mn}| \times |\mathcal{P}_{mn}|}$ and $Y_{mn}^s \in \mathbb{C}^{|\mathcal{P}_{mn}| \times |\mathcal{P}_{mn}|}$, respectively. Similarly, transformers are modeled as series components with an admittance matrix that depends on their connection type. Assembling the admittance matrices of distribution components, we construct the bus admittance matrix of the distribution system, denoted $Y_{\text{bus}} \in \mathbb{S}^{D \times D}$, which satisfies $Y_{\text{bus}} \mathbf{1} = \mathbf{0}$ if shunt elements are neglected. The bus admittance matrix relates the node voltages and injected currents according to Ohm's law:

$$\underbrace{\begin{bmatrix} I_1(k) \\ I_2(k) \\ \vdots \\ I_N(k) \end{bmatrix}}_{I_{\text{bus}}(k)} = \underbrace{\begin{bmatrix} Y_{11} & Y_{12} & \ldots & Y_{1N} \\ Y_{12}^\top & Y_{22} & \ldots & Y_{2N} \\ \vdots & \vdots & \ddots & \vdots \\ Y_{1N}^\top & Y_{2N}^\top & \ldots & Y_{NN} \end{bmatrix}}_{Y_{\text{bus}}} \underbrace{\begin{bmatrix} V_1(k) \\ V_2(k) \\ \vdots \\ V_N(k) \end{bmatrix}}_{V_{\text{bus}}(k)}, \quad (1)$$

where $V_{\text{bus}}(k), I_{\text{bus}}(k) \in \mathbb{C}^D$ are steady-state complex voltages and injected currents at time $k$, each off-diagonal block of $Y_{\text{bus}}$ is a submatrix $Y_{mn} = -Z_{mn}^{-1}$ corresponding to the admittance of line $(m, n)$, and each diagonal block is a submatrix

$$Y_{nn} = \sum_{m \in \{o|(o,n) \in \mathcal{E}\}} \left( \frac{1}{2} Y_{mn}^s + Z_{mn}^{-1} \right).$$

Rewriting (1) in vector form for time indices $k = 1, \ldots, K$ yields the following equation:

$$\underbrace{\begin{bmatrix} I_1(1) & \ldots & I_1(K) \\ I_2(1) & \ldots & I_2(K) \\ \vdots & \ddots & \vdots \\ I_N(1) & \ldots & I_N(K) \end{bmatrix}}_{I_{\text{bus}}^K} = Y_{\text{bus}} \underbrace{\begin{bmatrix} V_1(1) & \ldots & V_1(K) \\ V_2(1) & \ldots & V_2(K) \\ \vdots & \ddots & \vdots \\ V_N(1) & \ldots & V_N(K) \end{bmatrix}}_{V_{\text{bus}}^K}, \quad (2)$$

**Problem 1.** *IPF problem in three-phase distribution systems: Given steady-state measurements of voltage and current waveforms at different buses, $V_{bus}^K$ and $I_{bus}^K$, recover the bus admittance matrix $Y_{bus}$.*

We remark that the row rank of $V_{\text{bus}}^K$ is generally low in a distribution system[2]. As a result, the problem of solving (2) for $Y_{\text{bus}}$ (e.g. via ordinary least squares) is ill-posed in practice. In the following, we propose an identification algorithm that can deal with the low rank structure of voltage measurements.

---
[2]This phenomenon is reported for transmission PMU data in [8] and is supported by our experiments with realistic distribution PMU data [5].

**Algorithm 1** Basis Selection Algorithm

1: Perform orthogonal-triangular decomposition of $V_{\text{bus}}^K$;
2: Sort diagonal elements of the upper triangular matrix;
3: Choose the first $R$ that exceed a threshold and select the corresponding elements from the permutation matrix;
4: Return these elements as indices of linearly independent rows of $V_{\text{bus}}^K$;

### B. System Identification Algorithm

The standard least squares estimator fails to identify the bus admittance matrix due to the low-rank structure of $V_{\text{bus}}^K$. To address this problem, we propose an identification algorithm which exploits a particular partitioning of $V_{\text{bus}}^K$ into two matrices, one of which has full row rank.

*1) Similarity Transformation:* Let $R$ denote the row rank of $V_{\text{bus}}^K$. We partition $V_{\text{bus}}^K$ into two matrices through a similarity transformation:

$$\mathcal{T} I_{\text{bus}}^K = \underbrace{(\mathcal{T} Y_{\text{bus}} \mathcal{T}^{-1})}_{\mathbb{Y}} (\mathcal{T} V_{\text{bus}}^K), \quad (3)$$

where $\mathcal{T}$ is a $D \times D$ matrix that splits $V_{\text{bus}}^K$ into an $R \times K$ matrix, denoted $\mathbb{V}_2$, containing $R$ linearly independent rows of $V_{\text{bus}}^K$ and an $(D-R) \times K$ matrix, denoted $\mathbb{V}_1$, containing other rows of $V_{\text{bus}}^K$ that are all in the row space of $\mathbb{V}_2$. Algorithm 1 describes the steps for building these two submatrices from PMU data. Rearranging the rows of $V_{\text{bus}}^K$ and $I_{\text{bus}}^K$ according to this transformation thus yields:

$$\mathcal{T} V_{\text{bus}}^K = \begin{bmatrix} \mathbb{V}_1 \\ \mathbb{V}_2 \end{bmatrix}, \qquad \mathcal{T} I_{\text{bus}}^K = \begin{bmatrix} \mathbb{I}_1 \\ \mathbb{I}_2 \end{bmatrix}.$$

*2) Finding the Basis of $\mathbb{V}_1$:* Given this transformation, since $\mathbb{V}_1$ is in the row space of $\mathbb{V}_2$, we can write $\mathbb{V}_1 = X \mathbb{V}_2$. Hence, we can estimate the basis $X$ from PMU data by computing the pseudo-inverse of $\mathbb{V}_2$: $X = \mathbb{V}_1 \mathbb{V}_2^\dagger$. Note that the pseudo-inverse is well-defined here since $\mathbb{V}_2$ is full row rank.

*3) Estimating $\mathbb{Y}_X$:* We write (3) as

$$\begin{bmatrix} \mathbb{I}_1 \\ \mathbb{I}_2 \end{bmatrix} = \underbrace{\begin{bmatrix} \mathbb{Y}_{11} & \mathbb{Y}_{12} \\ \mathbb{Y}_{12}^\top & \mathbb{Y}_{22} \end{bmatrix}}_{\mathbb{Y}} \begin{bmatrix} X \mathbb{V}_2 \\ \mathbb{V}_2 \end{bmatrix} = \underbrace{\begin{bmatrix} \mathbb{Y}_{11} X + \mathbb{Y}_{12} \\ \mathbb{Y}_{12}^\top X + \mathbb{Y}_{22} \end{bmatrix}}_{\mathbb{Y}_X} \mathbb{V}_2 \quad (4)$$

Since $\mathbb{V}_2$ has full row rank, we can formulate the following least square problem which can be easily solved to determine the unique $\mathbb{Y}_X \in \mathbb{C}^{D \times R}$ from noisy measurements:

$$\mathbb{Y}_X = \arg \min_{\mathcal{Y} \in \mathbb{C}^{D \times R}} \left\| \begin{bmatrix} \mathbb{I}_1 \\ \mathbb{I}_2 \end{bmatrix} - \mathcal{Y} \mathbb{V}_2 \right\|_2. \quad (5)$$

*4) Recovering Components of $Y_{bus}$:* Once $\mathbb{Y}_X$ is identified, we focus on identifying matrices $\mathbb{Y}_{ij}$ for $i,j \in \{1,2\}$. We write out the equations:

$$\mathbb{I}_1 = (\mathbb{Y}_{11} X + \mathbb{Y}_{12}) \mathbb{V}_2 \quad (6)$$

$$\mathbb{I}_2 = (\mathbb{Y}_{12}^\top X + \mathbb{Y}_{22}) \mathbb{V}_2, \quad (7)$$

Solving (6) for $\mathbb{Y}_{12}$ and substituting it in (7) yields:

$$-X^\top * \mathbb{Y}_{11} * X + \mathbb{Y}_{22} = C \quad (8)$$

in which $C = \mathbb{I}_2 \mathbb{V}_2^\dagger - (\mathbb{V}_2^\dagger)^\top \mathbb{I}_1^\top X$ can be computed from PMU data. Applying the $\text{vec}(\cdot)$ operator to both sides of the equation, we obtain:

$$\left( X^\top \otimes X^\top \right) \text{vec}(\mathbb{Y}_{11}) + \text{vec}(\mathbb{Y}_{22}) = \text{vec}(C),$$

where $\otimes$ denotes the Kronecker product. Exploiting the sparsity of the admittance matrix of a (radial) distribution system, we now obtain $\mathbb{Y}_{11}$ and $\mathbb{Y}_{22}$ by solving the following optimization problem:

$$\min \left\| \begin{bmatrix} \text{vec}(\mathbb{Y}_{11}) \\ \text{vec}(\mathbb{Y}_{22}) \end{bmatrix} \right\|_0$$
$$\text{s.t.:} \begin{bmatrix} -X^\top \otimes X^\top & I \end{bmatrix} \begin{bmatrix} \text{vec}(\mathbb{Y}_{11}) \\ \text{vec}(\mathbb{Y}_{22}) \end{bmatrix} = \text{vec}(C), \quad (9)$$
$$\mathbb{Y}_{11} \in \mathbb{S}^{(D-R) \times (D-R)}, \mathbb{Y}_{22} \in \mathbb{S}^{R \times R}.$$

which can be relaxed to:

$$\min \left\| \begin{bmatrix} \text{vec}(\mathbb{Y}_{11}) \\ \text{vec}(\mathbb{Y}_{22}) \end{bmatrix} \right\|_1$$
$$\text{s.t.:} \begin{bmatrix} -X^\top \otimes X^\top & I \end{bmatrix} \begin{bmatrix} \text{vec}(\mathbb{Y}_{11}) \\ \text{vec}(\mathbb{Y}_{22}) \end{bmatrix} = \text{vec}(C), \quad (10)$$
$$\mathbb{Y}_{11} \in \mathbb{S}^{(D-R) \times (D-R)}, \mathbb{Y}_{22} \in \mathbb{S}^{R \times R}.$$

The above optimization problem is convex and can be solved efficiently. Once this problem is solved, we estimate $\mathbb{Y}_{12}$ from (6) using the method of least squares:

$$\mathbb{Y}_{12} = \arg \min_{\mathcal{Y} \in \mathbb{C}^{(D-R) \times R}} \left\| (\mathbb{Y}_{11} X + \mathcal{Y}) \mathbb{V}_2 - \mathbb{I}_1 \right\|_2 \quad (11)$$

Note that errors in estimating $\mathbb{Y}_{11}, \mathbb{Y}_{22}$ influence the estimation of $\mathbb{Y}_{12}$. In a separate line of work, we investigate the robustness of the algorithm to understand how errors propagate.

## IV. ONLINE EVENT DETECTION & LOCALIZATION

Many events, such as switching operations, tap changes, arc or ground faults, and other outages alter the admittance between certain nodes in a distribution network. Building on the IPF framework, in this section we design an efficient online algorithm for detecting and locating such events in a distribution system. The proposed algorithm requires only a small amount of data and has a low false alarm rate, enabling operators to take necessary remedial actions in quasi real-time.

Consider an affine parameterization of the admittance matrix, denoted $Y_{\text{bus}}^{\delta(k)}$, where $\delta(k) = \begin{cases} 0 & k < t \\ 1 & k \geq t \end{cases}$ is the discrete mode and $t$ is the time that the event has occurred. The proposed algorithm determines $t$ and finds out how the admittance matrix has changed by estimating the difference $Y_{\text{bus}}^1 - Y_{\text{bus}}^0$ using a small number of successive voltage and current phasor measurements. The changed entries of the admittance matrix indicate the type and the approximate location of the event.

### A. Detecting the Occurrence of an Event

To detect a change in the admittance matrix, we estimate the injected current vector at time $k$ from Ohm's law using the known admittance matrix, $Y_{\text{bus}}^0$, and the measured voltage vector at time $k$. We then compare the estimated injected current vector $\hat{I}_\text{bus}$ with the measured current vector $I_\text{bus}$ at time $k$ to calculate the prediction error:

$$e(k) = I_\text{bus}(k) - \hat{I}_\text{bus}(k) = I_\text{bus}(k) - Y_\text{bus}^0 V_\text{bus}(k), \quad (12)$$

The series $e(\cdot)$ is white noise if the admittance matrix does not change; this can be verified by the turning point test. When the prediction error $\|e(k)\|$ exceeds a predefined threshold $\gamma$, we assert that the admittance matrix has changed at $t = k$.

### B. Recovering the Admittance Matrix

After the occurrence of an event is detected, the admittance matrix must be recomputed to locate the event and update the network topology. Specifically, the difference between the admittance matrices of the system before and after the event will indicate the event type and can be used to pinpoint the event to a small number of possible locations. For example, if the difference between these matrices suggests that only two blocks corresponding to components installed between two distinct pairs of nodes have changed, it could be indicative of a switch that was opened while another one was closed.

A naive approach to event localization is therefore to rerun the identification algorithm presented in Section III-B upon detection of an event, and compare the inferred admittance matrix with the one identified prior to the detection. This requires processing at least $N$ successive PMU samples following the detection, implying that event identification in a system with many nodes cannot be accomplished shortly after the detection as enough PMU samples are not yet available. To address this shortcoming, we propose a promising identification algorithm that requires only a small number of PMU samples following the detection. This algorithm leverages the fact that only a few blocks of the admittance matrix will change due to an event; hence, the difference between the two admittance matrices, $Y_\text{bus}^1 - Y_\text{bus}^0$, is sparse. This allows us to formulate the following problem:

$$\min_{Y_\text{bus}^1} \|Y_\text{bus}^1 - Y_\text{bus}^0\|_0,$$
$$\text{s.t.:} \quad I_\text{bus}^{t \to t+K} = Y_\text{bus}^1 V_\text{bus}^{t \to t+K}, \quad Y_\text{bus}^1 \in \mathbb{S}^{D \times D} \quad (13)$$

in which $I_\text{bus}^{t \to t+K} = \left[I_\text{bus}(t), I_\text{bus}(t+1), \ldots, I_\text{bus}(t+K)\right]$, $V_\text{bus}^{t \to t+K} = \left[V_\text{bus}(t), V_\text{bus}(t+1), \ldots, V_\text{bus}(t+K)\right]$, and $t$ is the time slot when the event is detected. Let us define $\Delta Y \triangleq Y_\text{bus}^1 - Y_\text{bus}^0$. It can be readily seen that $\Delta Y$ is a symmetric complex matrix as it is the difference of two symmetric complex matrices. Hence, the optimization problem (13) can be written as:

$$\min \|\Delta Y\|_0,$$
$$\text{s.t.:} \quad I_\text{bus}^{t \to t+K} - Y_\text{bus}^0 V_\text{bus}^{t \to t+K} = \Delta Y V_\text{bus}^{t \to t+K} \quad (14)$$
$$\Delta Y \in \mathbb{S}^{D \times D}$$

which can be relaxed to the following $\ell 1$-norm optimization:

$$\min \|\Delta Y\|_1,$$
$$\text{s.t.:} \quad I_\text{bus}^{t \to t+K} - Y_\text{bus}^0 V_\text{bus}^{t \to t+K} = \Delta Y V_\text{bus}^{t \to t+K} \quad (15)$$
$$\Delta Y \in \mathbb{S}^{D \times D}$$

This is a convex problem and can be efficiently solved. More importantly, solving this problem requires only a small number of PMU samples unlike the naive approach.

## V. EXPERIMENTS

To evaluate the performance of the proposed event detection and system identification algorithms we run power flow analysis on the IEEE 13 bus test feeder [23] using the Open Distribution System Simulator (OpenDSS) [24]. The proposed algorithm is implemented in MATLAB and the optimization problems are solved using the CVX toolbox [25]. This section describes our simulation scenarios and presents the results for inferring admittance matrix and detecting a line tripping event.

### A. Simulation Scenarios

The IEEE 13-bus feeder is a three-phase, unbalanced radial distribution system. The OpenDSS model of this network contains 38 nodes, which is inclusive of equipment terminals and different phases of its 13 buses. We assume that a distribution PMU is installed at every node and treat nodes as load aggregation points. Thus, the load connected to each node represents the aggregated demand of a certain number of downstream households, where the demand of every household is generated by continuous time Markov models derived from realistic residential loads in [26]. The synthesized aggregated demands are sampled at 120Hz to simulate phasor measurements [11]. We assume that all nodes consume real and reactive power except those corresponding to the source bus, the voltage regulator terminals, and buses 650 and 692. In our simulations, the distribution of loads across different phases of a node is nonuniform and a constant power factor of 95% is assumed at each node. We connect a total of 3300 households with peak-to-average ratios between 1.47 and 1.20 to this radial system as described in Table I.

Our simulations span over one day divided into time slots of equal length, each taking one half of the AC cycle. We update the demand of all nodes in every time slot and perform power flow calculations subsequently. The OpenDSS simulator returns complex voltages and currents injected at all nodes, which are treated as phasor measurements for that time slot.

### B. Event Detection and Localization Results

We first verify that the algorithm presented in Section III-B is capable of inferring the admittance matrix of this radial system from the PMU data under normal operation. We consider the error of estimating $\mathbb{Y}_{22}$ as our performance metric. Our simulations show that all elements of $\mathbb{Y}_{22}$ can be estimated with above 98.5% accuracy.

We next evaluate our event detection and localization algorithm. To this end, we introduce a line tripping event by disconnecting the single-phase line between buses 611 and 684. This event will change the admittance matrix that can be inferred from data. We assume that the system admittance matrix has been identified with high accuracy prior to this event, as described above. We observe that the proposed algorithm detects the event in the same time slot that it occurs, i.e., after processing the PMU data for that time slot only.

TABLE I
NUMBER OF HOMES CONNECTED TO THE PHASES OF EACH NODE (B: BUS IDENTIFIER, P: PHASE IDENTIFIER, H: NUMBER OF HOMES).

| B | 632 | | | 671 | | | 680 | | | 633 | | | 634 | | | 675 | | | 645 | | 646 | | 684 | | 652 | 611 |
|---|---|---|---|---|---|---|---|---|---|---|---|---|---|---|---|---|---|---|---|---|---|---|---|---|---|---|
| P | a | b | c | a | b | c | a | b | c | a | b | c | a | b | c | a | b | c | b | c | b | c | a | c | a | c |
| H | 300 | 280 | 250 | 10 | 15 | 25 | 130 | 180 | 200 | 40 | 50 | 60 | 60 | 50 | 40 | 125 | 100 | 75 | 70 | 30 | 190 | 210 | 55 | 45 | 150 | 150 |

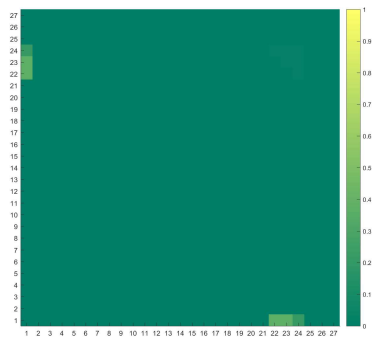

Fig. 1. The identification error $|\mathbb{Y}_{22}^1 - \hat{\mathbb{Y}}_{22}^1|$.

Moreover, it successfully estimates an important part of the new admittance matrix, i.e., 27 nodes out of the 39 nodes whose voltage measurements are linearly independent. Let $\hat{\mathbb{Y}}_{22}^1$ and $\mathbb{Y}_{22}^1$ denote the inferred and the true admittance submatrix pertaining to these 27 nodes, respectively. Figure 1 shows the identification error defined as $|\mathbb{Y}_{22}^1 - \hat{\mathbb{Y}}_{22}^1|$. The color of a cell located at row $i$ and column $j$ represents the value of $|\mathbb{Y}_{22}^1(i,j) - \hat{\mathbb{Y}}_{22}^1(i,j)|$ as shown in the color bar. It can be seen that the identification error is relatively small compared to the absolute value of the elements of $\mathbb{Y}_{22}^1$ (below 0.4%). Furthermore, comparing $\hat{\mathbb{Y}}_{22}^1$ with $\hat{\mathbb{Y}}_{22}^0$ indicates that the admittance of the single-phase line connecting buses 611 and 684 has changed due to this event, enabling us to locate the event within a small geographical area.

## VI. CONCLUSIONS

Distribution system topology detection, and event detection and localization are crucial steps towards enhancing the resilience of modern distribution grids. The availability of high-fidelity, high-sample-rate measurements from various locations in a distribution system has offered a tremendous potential for accomplishing these tasks within the timescale of power system operation. Building on the inverse power flow framework, we design an online algorithm which is capable of detecting and locating critical distribution system events within a small geographical area using only a small number of successive voltage and current phasor measurements, enabling system operators to initiate remedial actions in a timely manner. Simulations performed on a test distribution feeder corroborate the effectiveness of this algorithm despite the low rank structure of the PMU data. Future efforts will develop theoretical underpinnings and further investigate the proposed method in terms of its performance and sensitivity to measurement noise.

## REFERENCES

[1] G. Kaur and M. Vaziri, "Effects of distributed generation (DG) interconnections on protection of distribution feeders," in *IEEE PES General Meeting*, 2006.
[2] N. Seltenrich, "The New Grid - Plugging into California's clean-energy future," 2013. [Online]. Available: https://nature.berkeley.edu/breakthroughs/fa12/the_new_grid
[3] R. C. Green, L. Wang, and M. Alam, "The impact of plug-in hybrid electric vehicles on distribution networks: A review and outlook," *Renewable and Sustainable Energy Reviews*, vol. 15, no. 1, pp. 544–553, 2011.
[4] E. Veldman and R. Verzijlbergh, "Distribution Grid Impacts of Smart Electric Vehicle Charging From Different Perspectives," *IEEE Transactions on Smart Grid*, vol. 6, no. 1, pp. 333–342, Jan. 2015.
[5] A. von Meier, D. Culler, A. McEachern, and R. Arghandeh, "Micro-synchrophasors for distribution systems," in *IEEE PES Innovative Smart Grid Technologies Conference*, Feb 2014, pp. 1–5.
[6] North American Synchrophasor Initiative, "Distribution Task Team," https://www.naspi.org/distt.
[7] R. Dobbe, D. Arnold, S. Liu, D. Callaway, and C. Tomlin, "Real-time distribution grid state estimation with limited sensors and load forecasting," in *ACM/IEEE 7th International Conference on Cyber-Physical Systems*, 2016, pp. 1–10.
[8] L. Xie, Y. Chen, and P. R. Kumar, "Dimensionality Reduction of Synchrophasor Data for Early Event Detection: Linearized Analysis," *IEEE Transactions on Power Systems*, vol. 29, no. 6, pp. 2784–2794, Nov 2014.
[9] A. Pal, P. Chatterjee, J. S. Thorp, and V. A. Centeno, "Online calibration of voltage transformers using synchrophasor measurements," *IEEE Transactions on Power Delivery*, vol. 31, no. 1, pp. 370–380, Feb 2016.
[10] Y. Yuan, O. Ardakanian, S. Low, and C. Tomlin, "On the inverse power flow problem," https://arxiv.org/abs/1610.06631, Tech. Rep., Oct. 2016.
[11] PSL, "Micro Synchrophasors," http://www.powersensorsltd.com/PQube3.php.
[12] X. Li, H. V. Poor, and A. Scaglione, "Blind topology identification for power systems," in *Smart Grid Communications, IEEE International Conference on*, Oct 2013, pp. 91–96.
[13] Y. Liao, Y. Weng, M. Wu, and R. Rajagopal, "Distribution grid topology reconstruction: An information theoretic approach," in *North American Power Symposium*, Oct 2015, pp. 1–6.
[14] D. Deka, S. Backhaus, and M. Chertkov, "Estimating distribution grid topologies: A graphical learning based approach," *arXiv preprint arXiv:1602.08509*, 2016.
[15] Q. Huang, L. Shao, and N. Li, "Dynamic Detection of Transmission Line Outages using Hidden Markov Models," in *American Control Conference (ACC)*. IEEE, 2015, pp. 5050–5055.
[16] J. E. Tate and T. J. Overbye, "Line Outage Detection Using Phasor Angle Measurements," *IEEE Transactions on Power Systems*, vol. 23, no. 4, pp. 1644–1652, Nov 2008.
[17] W. Pan, Y. Yuan, H. Sandberg, J. Goncalves, and G.-B. Stan, "Online fault diagnosis for nonlinear power systems," *Automatica*, vol. 55, pp. 27–36, 2015.
[18] Y. C. Chen, T. Banerjee, A. D. Dominguez-Garcia, and V. V. Veeravalli, "Quickest Line Outage Detection and Identification," *IEEE Transactions on Power Systems*, vol. 31, no. 1, pp. 749–758, Jan 2016.
[19] H. Zhu and G. B. Giannakis, "Sparse Overcomplete Representations for Efficient Identification of Power Line Outages," *IEEE Transactions on Power Systems*, vol. 27, no. 4, pp. 2215–2224, Nov 2012.
[20] R. Emami and A. Abur, "External system line outage identification using phasor measurement units," *IEEE Transactions on Power Systems*, vol. 28, no. 2, pp. 1035–1040, May 2013.
[21] Y. Sharon, A. M. Annaswamy, A. L. Motto, and A. Chakraborty, "Topology identification in distribution network with limited measurements," in *IEEE PES Innovative Smart Grid Technologies*, Jan 2012, pp. 1–6.
[22] G. Cavraro, R. Arghandeh, K. Poolla, and A. von Meier, "Data-driven approach for distribution network topology detection," in *IEEE PES General Meeting*, July 2015, pp. 1–5.
[23] W. Kersting, "Radial distribution test feeders," in *IEEE PES Winter Meeting*, vol. 2, 2001, pp. 908–912.
[24] EPRI, "Simulation Tool: OpenDSS," http://www.smartgrid.epri.com/SimulationTool.aspx.
[25] M. Grant and S. Boyd, "CVX: Matlab software for disciplined convex programming, version 2.1," http://cvxr.com/cvx, Mar. 2014.
[26] O. Ardakanian, S. Keshav, and C. Rosenberg, "Markovian Models for Home Electricity Consumption," in *SIGCOMM Workshop on Green Networking*. ACM, 2011, pp. 31–36.